\documentclass[aps,twocolumn,groupedaddress,showkeys,floatfix,superscriptaddress,10pt]{revtex4-2}

\usepackage{graphicx}
\usepackage{epsfig}
\usepackage[T1]{fontenc}

\usepackage{amssymb}
\usepackage{amsmath}
\usepackage{color}
\usepackage{float}
\usepackage{xcolor}
\usepackage[utf8]{inputenc}
\usepackage[english]{babel}

\usepackage{sidecap}
\sidecaptionvpos{figure}{c}
\renewcommand\thesection{\Roman{section}}

\begin{document}

\title{VO$_2$ Nanocrystals for Designer Phase-Change Metamaterials}

\author{Jimmy John}
\affiliation{Institut des Nanotechnologies de Lyon, Ecole Centrale de Lyon, 69134 Ecully, France}

\author{Yael Gutierrez}
\affiliation{Department of Applied Physics, Universidad de Cantabria, Avda. Los Castros s/n 39005 Santander, Spain}
\author{Zhen Zhang}
\affiliation{School of Materials Engineering, Purdue University, West Lafayette, Indiana 47907, USA}
\author{Helmut Karl}
\affiliation{Lehrstuhl für Experimentalphysik IV, Universität Augsburg, 86159 Augsburg, Germany}
\author{Shriram Ramanathan}
\affiliation{School of Materials Engineering, Purdue University, West Lafayette, Indiana 47907, USA}
\author{R\'{e}gis Orobtchouk}
\affiliation{Institut des Nanotechnologies de Lyon, Ecole Centrale de Lyon, 69134 Ecully, France}
\author{Fernando Moreno}
\affiliation{Department of Applied Physics, Universidad de Cantabria, Avda. Los Castros s/n 39005 Santander, Spain}

\author{S\'{e}bastien Cueff}
\affiliation{Institut des Nanotechnologies de Lyon, Ecole Centrale de Lyon, 69134 Ecully, France}

\date{\today}

\begin{abstract}
Subwavelength nanoparticles can support electromagnetic resonances with distinct features depending on their size, shape and nature. For example, electric and magnetic Mie resonances occur in dielectric particles, while plasmonic resonances appear in metals. Here, we experimentally demonstrate that the multipolar resonances hosted by VO$_2$ nanocrystals can be dynamically tuned and switched thanks to the insulator-to-metal transition of VO$_2$. Using both Mie theory and Maxwell Garnett effective medium theory, we retrieve the complex refractive index of the effective medium composed of a slab of VO$_2$ nanospheres embedded in SiO$_2$ and show that such a resulting metamaterial presents distinct optical tunability compared to unpatterned VO$_2$. We further show that this provides a new degree of freedom to design low-loss phase-change metamaterials with designer optical tunability and actively controlled light scattering.
\end{abstract}

\maketitle
\vspace{-0.4cm}
\section{Introduction}
\vspace{-0.4cm}
The potential of nanophotonics for tailoring light-matter interaction at the nanoscale has attracted considerable interest in recent years~\cite{novotny2012principles,koenderink2015nanophotonics, kivshar2017meta}. For example, metasurfaces enable an ultimate control of light fields thanks to abrupt phase modifications using engineered nanoscale elements~\cite{chen,yu2014flat}. A wealth of nano-fabrication techniques are readily available to produce nanoscale elements with specific shapes, sizes and nature. Thanks to these technological developments, light scattering by spherical nanoparticles revealed exciting optical phenomena such as strong localized optical resonances, directional scattering or light emission control \cite{Kelly2003, Garcia-Etxarri2011, Geffrin2012, bidault2019dielectric}.
However, the dimensions of these nano-resonators, and the scale of nanophotonic devices make them difficult to be tuned and reconfigured. So in essence most of nanophotonic devices are static devices. 
It is therefore currently a great challenge to find efficient means to dynamically tune photonic devices at the nanoscale~\cite{ferrera2017dynamic,zheludev2015obtaining}. 

Recently, a large number of researches make use of phase-change materials (PCM) to tune photonic devices. Although a large variety of PCMs exist, most of the works exploit the tunable properties of either VO$_2$ or GeSbTe~\cite{wuttig2017phase,ke2018vanadium,shi2019recent,cueff2015dynamic}. In the standard visible (Vis) and near-infrared (NIR) ranges, these materials typically have large refractive index modulation but also large extinction coefficients. So far, it has proven difficult to circumvent this extinction coefficient, which translates into net optical losses in photonic devices.
One interesting approach to modify the intrinsic properties of these materials is to locally arrange or modify them by patterning, doping or straining. These strategies were recently used to demonstrate phase modulation~\cite{kim}, tunable optical absorption~\cite{zhu,rensberg} or switchable dielectric-plasmonic regimes~\cite{butakov}. A particularly promising method is to reduce the dimensions of VO$_2$ down to the nanoscale. This approach is not only interesting for gaining theoretical insights into the fundamentals of the Insulator-to-Metal Transition (IMT) of VO$_2$~\cite{whittaker2011microscopic} but also to tailor the physical properties of this strongly correlated system. 
The first works on synthesis and characterization of VO$_2$ nanocrystals (NCs) were reported in 2002~\cite{lopez2002synthesis,lopez2002size}, followed by studies on their nonlinear optical properties~\cite{lopez2004optical} and their potential for ultra-fast modulation of optical transmission~\cite{rini}.
Recent works report various methods of fabrication of VO$_2$ nanocrystals and their use as thermochromic smart windows~\cite{wei2019,zhuang2019}, differentially-doped multilayer VO$_2$ films~\cite{paik}, electrochemically-induced transformations~\cite{dahlman2016electrochemically} or dynamical reconfiguration of optical devices~\cite{zimmer,jostmeier,jostmeier2016}.

Here, we show that an ensemble of VO$_2$ nanospheres support multipolar resonances and can be homogenized as a tunable effective medium metamaterial whose optical properties and tunability are adjustable by design. Specifically, we exploit a Mie theory-based extension of Maxwell-Garnett effective medium approximation to precisely explain the light-matter interactions at play in VO$_2$-NCs and decompose it into multipolar modes. These multipolar resonances are both dynamically tunable through the IMT of VO$_2$ and adjustable via controlling the size of the particles. VO$_2$-NCs is therefore a tunable metamaterial platform whose properties can both be engineered through fabrication (size and density of VO$_2$-NCs) and actively tunable by external excitations (switching from insulator to metal).

\vspace{-0.4cm}
\section{Optical properties of VO$_2$ thin-films}
\vspace{-0.4cm}
VO$_2$ is a strongly correlated material that is dielectric at room temperature and becomes metallic when heated above $68^\circ$C. According to band theory and given its crystallography, it should be metallic even at room temperature, but electron-electron correlations freeze the potential free carriers on their respective sites (Mott localisation)~\cite{zylbersztejn1975metal,RevModPhys.40.677,imada1998metal}. Upon the transition, several interrelated physical processes occur: the crystallographic lattice changes from monoclinic to tetragonal (Peierls distortion), hence modifying the band structure and “liberating” free carriers. VO$_2$ therefore undergoes large changes in its electrical properties and its band structure, what translates into large modulation of its optical properties. 

\begin{figure}[H]
  \centering  \includegraphics[width=0.90\linewidth]{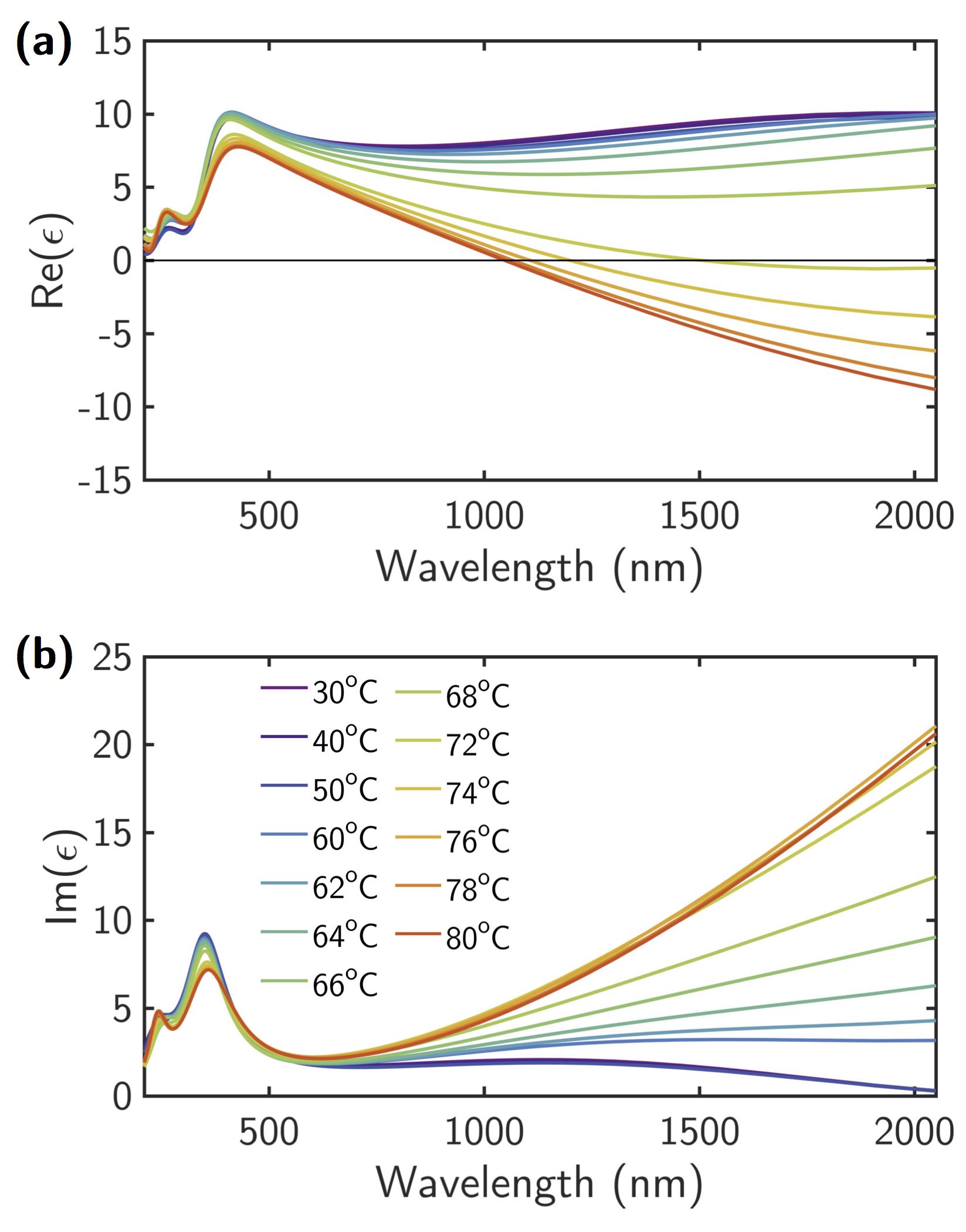}
  \caption{Real and imaginary part of the dielectric permittivity $\epsilon$ of a VO$_2$ thin-film in the Vis/NIR range as a function of temperature.}
  \label{fig1}
\end{figure}

We have investigated the optical properties of a thin layer of VO$_2$ across the IMT using spectroscopic ellipsometry combined with a heat cell (more details in the experimental section and supplementary information). Thanks to a physically consistent optical model (more details in the supplementary), we have extracted the dielectric function of VO$_2$ as a function of temperature. In Figure \ref{fig1}, we show the complex optical permittivity of VO$_2$ upon phase-change. As displayed in the figure, we see very large modifications of both the real (${\epsilon }_1$) and imaginary parts (${\epsilon }_2$) of the permittivity upon the IMT of VO$_2$. 

The extraordinary feature of VO$_2$ is the strong modulation of its dielectric permittivity, which is as large as $\Delta\epsilon\sim$ 20 in the near-infrared range. But, as explained previously, this modulation is accompanied by an intrinsic very large increase of the optical absorption, which dominates the optical response in the infrared range~\cite{wan2019optical,sun2017analyzing}. Given this large absorption, VO$_2$ is therefore often used as an on-off switch and little to no works actually make use of its refractive index tunability.

\vspace{-0.4cm}
\section{VO$_2$ nanocrystals}
\vspace{-0.4cm}
When an object is illuminated by an electromagnetic wave, it reradiates parts of the energy while being lost from the original wave. This phenomenon is known as scattering. The quantity and direction of light that is scattered depend on the nature, shape and size of the object \cite{Noguez2007}. The problem becomes substantially more complex when the size of the object approaches the wavelength of the incident wave. In that situation, we can no longer assume the field to be constant in the object and the so-called “long-wavelength approximation” no longer holds, so retardation effects must be considered. To compute the scattered field, one needs a numerical approach for an arbitrary object. Such a complex problem can however be simplified if the object is a sphere. In 1908, Mie found an exact solution to that problem, by calculating the scattered field as a series solution \cite{Mie1908}. The general idea is to expand the electric and magnetic fields in vector spherical harmonics. The field functions are then linear combinations of terms that are products of separable functions of the three spherical coordinates.
The electromagnetic interaction of light with spheres can then be modeled following the Lorenz-Mie formalism for scattering and absorption of light by small particles~\cite{bohren}. Within this formalism, the extinction, scattering and absorption efficiencies are given by \\

	\begin{equation}
	Q_{ext}=\frac{2}{x^2}\sum_{n=1}^{\infty}(2n+1)(|a_{n}|^2+|b_{n}|^2)
	\end{equation}
	
	\begin{equation}
	Q_{sca}=\frac{2}{x^2}\sum_{n=1}^{\infty}(2n+1)Re(a_{n}+b_{n})
	\end{equation}
	
	\begin{equation}
	Q_{abs}=Q_{ext}-Q_{sca}
	\end{equation}

\noindent where $a_n$ and $b_n$ are the so-called scattering coefficients~\cite{bohren}. These depend on both the particle optical properties (relative to its surrounding medium) and size. The size parameter $x$ is defined as  $x = \frac{2\pi m_{med} R}{\lambda}$ where $m_{med}$ is the refractive index of the surrounding medium, $\lambda$ the wavelength of the incident light in vacuum and $R$ the sphere's radius. Physically, $a_n$ and $b_n$ are the weighting factors of the different excited electric and magnetic multipolar contributions. For instance, $a_1$ and $b_1$ represent the electric and magnetic dipolar modes, and $a_2$ and $b_2$ correspond to the electric and magnetic quadrupolar modes, respectively.

\begin{figure}[H]
\begin{center}
\includegraphics[width=8.5cm]{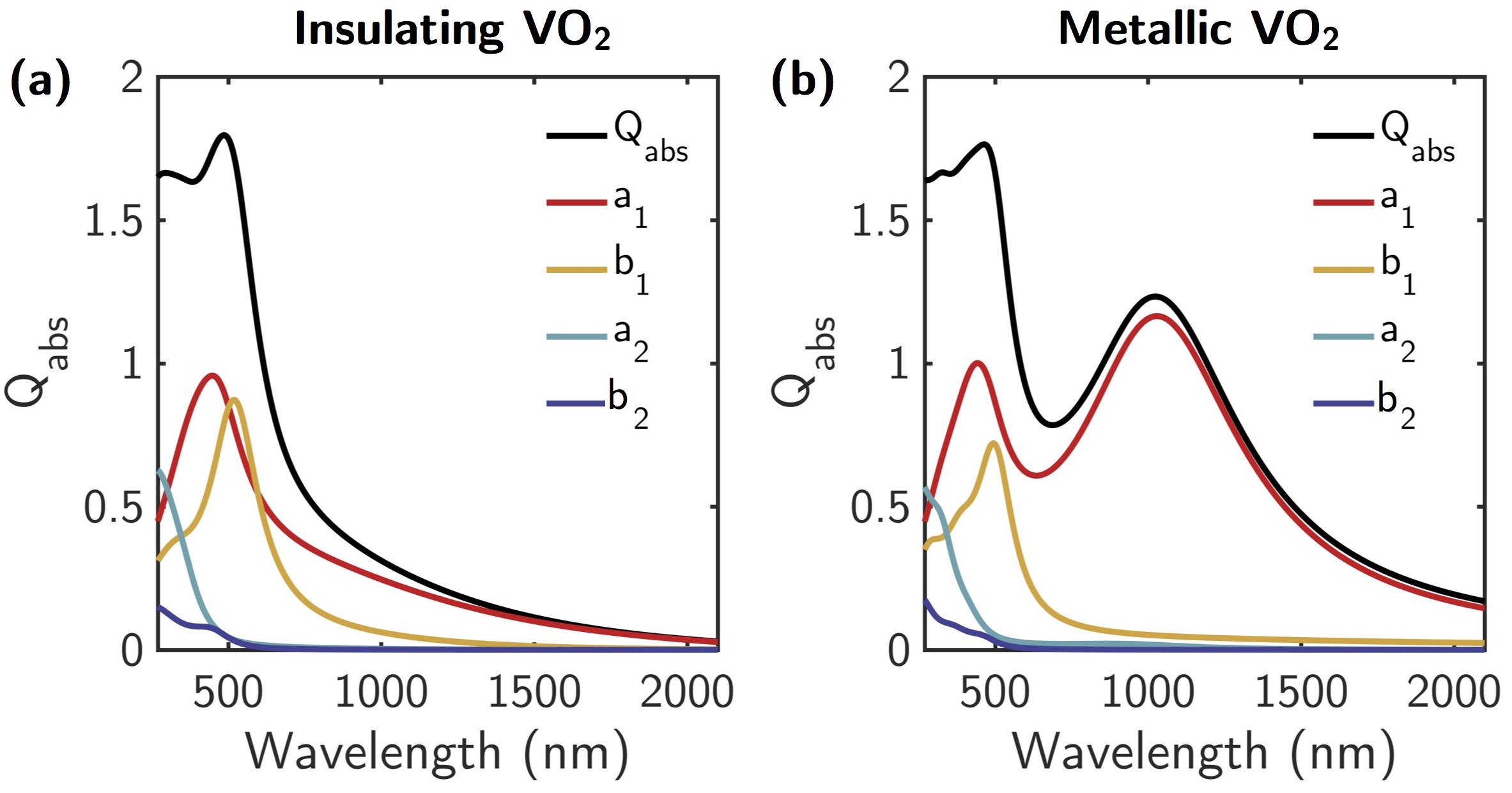}
\end{center}
\caption{{Calculated absorption efficiency of VO$_2$ spheres with radius 50 nm in its a) insulating and b) metallic states embedded in quartz. With colored lines are represented the dipolar electric $a_1$ (red), dipolar magnetic $b_1$ (yellow), quadrupolar electric $a_2$ (light blue) and quadrupolar magnetic $b_2$ (dark blue) contributions to $Q_{abs}$.}}
  \label{fig2}
\end{figure}

Figure \ref{fig2} shows the absorption cross section $Q_{abs}$ of a VO$_2$ sphere of radius 50 nm in its insulating and metallic states embedded in quartz ( $m_{med}$ = 1.45 for $\lambda = 1$ $\mu$m). With colored lines, dipolar electric $a_1$, dipolar magnetic $b_1$, quadrupolar electric $a_2$ and quadrupolar magnetic $b_2$ contributions to $Q_{abs}$ are represented. It can be seen that, regardless of the state of VO$_2$, below 700 nm the electromagnetic response is dominated by the dipolar electric (DE) and magnetic response (DM), with a small contribution of the quadrupolar electric term (QE). These resonances correspond to Whispering Gallery Modes (WGM) which resonate at the suitable wavelengths, the magnetic resonance redshifted with respect to the electric one \cite{Geffrin2012}. On the other hand, when VO$_2$ is in its metallic phase, a DE resonance appears at a longer wavelength (1015 nm). The higher the temperature and consequently, with increasing metallic character, the more intense this resonance is. This resonance is plasmonic and its physical origin is the negative value of the dielectric constant above 1000 nm in the metallic phase.

\begin{figure*}[tp]
\centering
\includegraphics[width=0.85\linewidth]{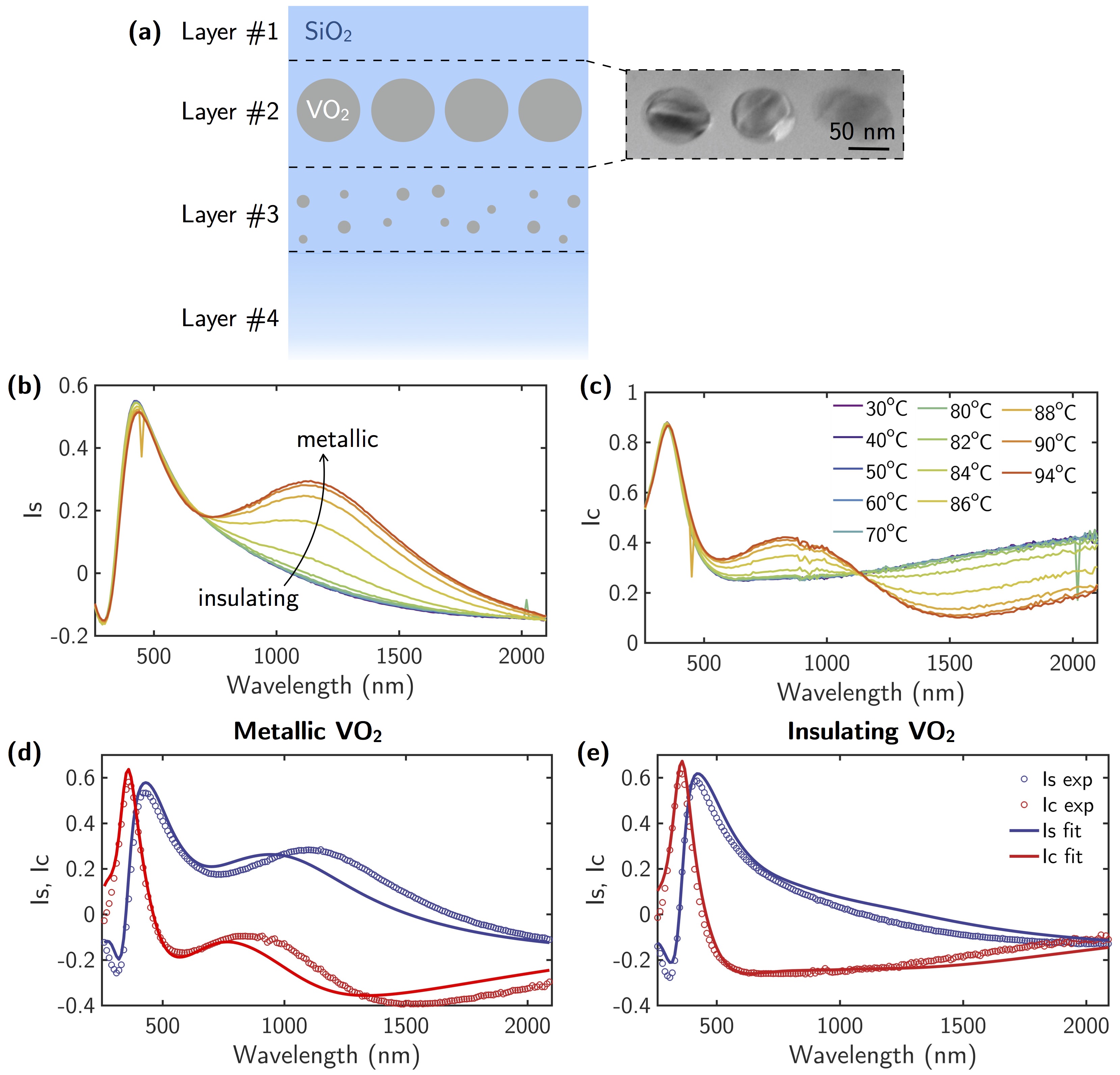}
\caption{{a) Sketch representing a cross-sectional view of the VO$_2$-NCs implanted in SiO$_2$ together with a TEM image of the VO$_2$-NCs layer. b) and c) Spectroscopic ellipsometry measurements of VO$_2$-NCs b) $Is$ parameter c) $Ic$ parameter, both measured as a function of temperature (incident angle: 65$^\circ$). d) and e) Fitting (solid line) and experimental measurements (circles) of the ellipsometric parameters $I_s$ (blue) and  $I_c$ (red) for both states of VO$_2$-NCs: d) metallic and e) insulating. (incident angle of 55$^\circ$ is displayed here for a better clarity of the superimposed $Is$ and $Ic$ parameters.)}}
\label{fig3}
\end{figure*}

An ensemble of VO$_2$ nanospheres would therefore enable a dynamic tuning of multipolar resonances together with their coherent effects for directionality purposes \cite{Tribelsky2015} and to actively control the presence and intensity of a plasmonic mode in the NIR.

\begin{figure*}[tb]
  \centering
  \includegraphics[width=\linewidth]{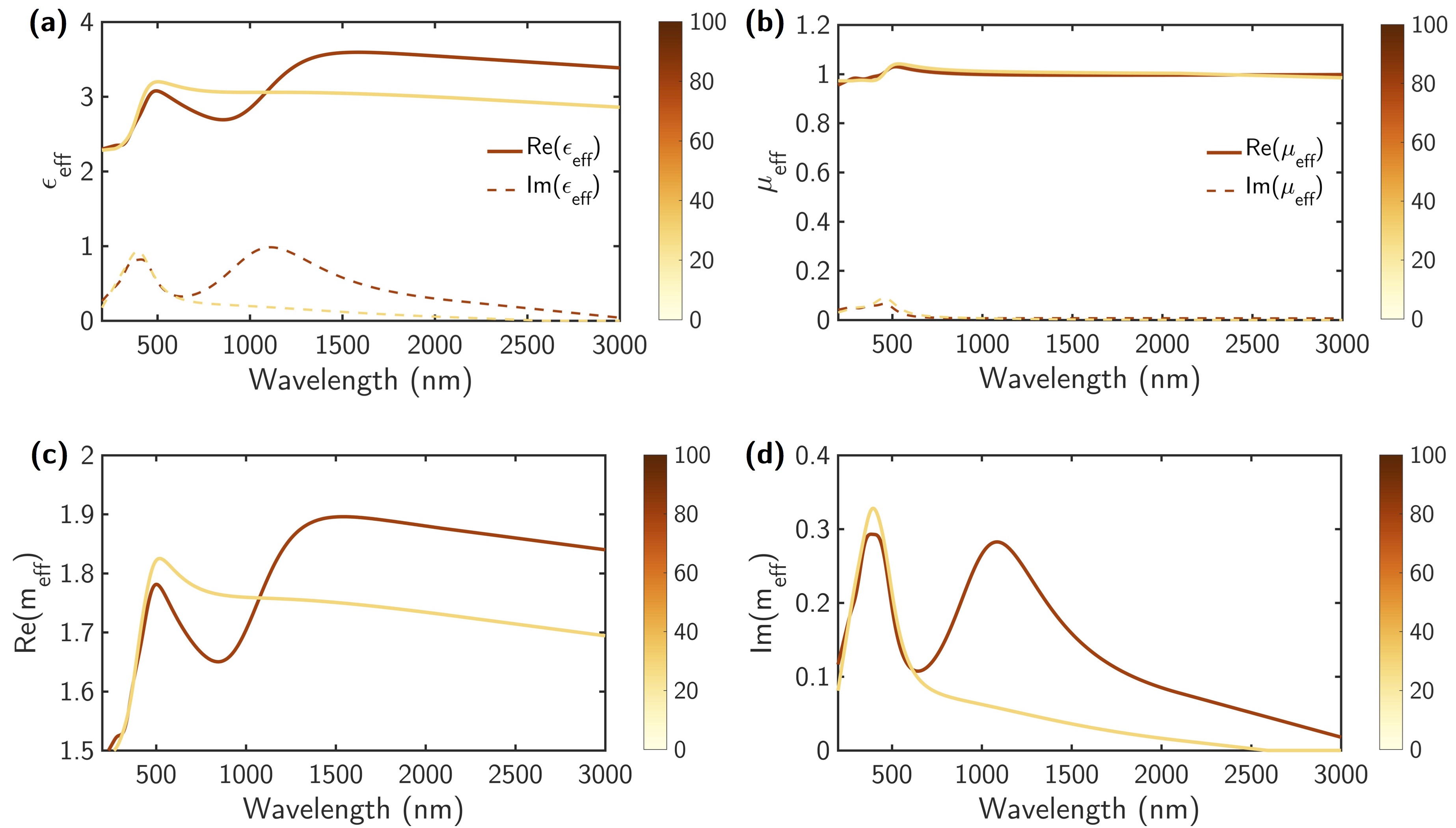}
  \caption{a) Effective dielectric function ($\epsilon_{eff}$), b) magnetic permeability (${\mu_{eff}}$), and c) real and d) imaginary parts of the effective refractive index ($m_{eff} = n_{eff} + ik_{eff}$) of a slab of quartz with VO$_2$ spheres with radius $R =$ 35 nm and a filling fraction $f =$ 0.18 in their insulating (low temperature) and metallic (high temperature) states.}
  \label{fig4}
\end{figure*}

\vspace{-0.4cm}
\section{Fabrication and optical characterization}
\vspace{-0.4cm}

We have measured and analyzed VO$_2$-NCs fabricated in a fused silica host co-implanted with vanadium ions and oxygen ions followed by a rapid thermal annealing process (more details in the experimental section).
After these steps, spherical VO$_2$-NCs are formed at some distance from the surface (for more details, see e.g.~\cite{jostmeier}). 
As verified by TEM analysis (see Fig. \ref{fig3}a), the produced sample contains an ensemble of VO$_2$ nanospheres that are close to be perfectly spherical~\cite{jostmeier2016}. 

To analyze the optical properties of VO$_2$-NCs, we used a spectroscopic ellipsometer (UVISEL plus Horiba), from 250 nm to 2100 nm, coupled to a heating cell to thermally induce the IMT (more details in the experimental section and in the supplementary information). 
Figure \ref{fig3}b-c show the evolution of the $I_s$ and $I_c$ parameters of ellipsometry measurements as a function of temperature, the effect of the IMT of VO$_2$-NCs is clearly observable on the raw ellipsometry measurements, with the appearance of a large peak in $I_s$ whose intensity progressively increases around 1100 nm. This phenomenon is reversible, with a broad hysteresis behavior (more details in the supplementary information), and is a clear signature of the presence of stoichiometric VO$_2$ in the sample.

\vspace{-0.4cm}
\section{Retrieval of the effective parameters of the metamaterial.}
\vspace{-0.4cm}

We now want to retrieve the effective permittivity of the active layer containing VO$_2$-NCs. From previous TEM analysis~\cite{jostmeier2016}, we know that the VO$_2$-NCs medium is a single-plane of NCs that are spaced by a few tens of nanometers. 
In order to model the optical response of the slab composed of VO$_2$-NCs embedded in quartz, we have calculated its effective refractive index m$_{eff}$ using a Mie theory-based extension of Maxwell Garnett effective medium approximation (EMA), as proposed by Doyle~\cite{ruppin}. Using this approach, we can calculate an effective dielectric permittivity (${\epsilon_{eff}}$) and magnetic permeability (${\mu_{eff}}$) that are directly governed by the Mie dipolar electric $a_1$ and magnetic $b_1$ dipolar coefficients respectively. In this way, we take into account size effects that are ignored by the electrostatic approximation which is the base of conventional Maxwell Garnett EMA. The dipolar coefficients are enough to model the response of the VO$_2$ particles since the quadrupolar terms are negligible as seen in Figure 1. Therefore, ${\epsilon_{eff}}$ and ${\mu_{eff}}$ can be written as

	\begin{equation}
	{\epsilon}_{eff}=\frac{x^3+3ifa_1}{x^3-\frac{3}{2}ifa_1}
	\end{equation}
	
	\begin{equation}
	{\mu}_{{eff}}=\frac{x^3+3ifb_1}{x^3-\frac{3}{2}ifb_1}
	\end{equation}	

where $x$ is the size parameter of the VO$_2$ spheres and $f$ the volume filling fraction of the VO$_2$ spheres in the slab (the complete derivation for eqs. (4) and (5) can be found in the supplementary). From the values of ${\epsilon_{eff}}$ and ${\mu_{eff}}$ the effective complex refractive index of the slab can be calculated as

\begin{equation}
	m_{eff}= n_{eff} + i k_{eff}=\sqrt{\epsilon_{eff}.\mu_{eff}}
\end{equation}

In the following, we fit the experimental ellipsometry measurements with a multi-layer model composed of four layers as sketched in Figure \ref{fig3}a: (i) a thin SiO$_2$ layer, (ii) the VO$_2$-NCs + SiO$_2$ effective layer, (iii) an intermediate layer containing vanadium inclusions and impurities and (iv) a semi-infinite SiO$_2$ substrate. This four-layer model is justified by TEM observations of the sample’s cross sections. We use a reference dispersion file for SiO$_2$ (Palik \cite{Palik1998}).
The respective thicknesses of SiO$_2$ top layer, VO$_2$-NC layer and V inclusion layer are set as free fit parameters. None of the material’s dispersion are fit, rather we modify the size and density of VO$_2$-NCs in our Mie-Maxwell Garnett model to adjust the fits to the measurements. As there could exist correlations between thicknesses and permittivity in absorbing films, we simultaneously fit the two sets of measurements for VO$_2$ insulating and VO$_2$ metallic and we bind the different thicknesses. In other words, the final fit will yield the same thicknesses for the different layers whatever the phase of VO$_2$-NCs. The only difference between the two models lies in the dispersion of VO$_2$-NCs.
In Figure \ref{fig3}d-e, we show the optimized fits together with the experimental data. We obtain an overall good correspondence between model and measurements for a VO$_2$-NCs size of $R=35nm$ and a filling-factor $f=0.18$. Especially, the fits are nearly perfect in the UV/Vis range but a slight discrepancy appears in the NIR.

The observed discrepancy in \ref{fig3}d-e between the model and the experimental results is due to different effects which cannot be included without losing simplicity in the proposed model. The two most important are the electromagnetic interaction between the particles and a potentially non-negligible degree of polydispersity in the size distribution of the VO$_2$-NCs. The electromagnetic interaction between nanoparticles has a long standing history and a nice work on this effect can be found in Rechberger et al.~\cite{rechberger2003optical}. We have applied the results of this research to our case and the details can be found in the supplementary information. Here, we have considered the experimental conditions used in the ellipsometry measurements with an exciting beam linearly polarized at 45º with respect to the incidence plane. In the metallic phase, the pure plasmonic resonance is affected by blue and red shifts due to the interparticle electromagnetic interaction with a clear net red shift of this resonance. This would explain the apparent discrepancy observed between the measured spectral position of the plasmonic peak around 1000 nm and that predicted with our model where multiple interactions are not included. Concerning the insulating phase, its resonance is clearly less influenced by electromagnetic interaction between the particles. It is important to point out that this resonance is composed mainly of two dipolar ones (see Figure 1) of electric and magnetic character, respectively. As it is shown in the supplementary information, due to the particle interaction, there is a competition between the blue shift undergone by the magnetic contribution and the red shift for the electric one~\cite{Albella2013}. The net result is a slight blue shift of the whole resonance when the interparticle gap is small. This is the small discrepancy we observe in Figure 3 when the experiment and the model result are compared for the resonance at 380 nm.

Concerning polydispersity, some detailed calculations have been included in the supplementary information but the main conclusion is that, assuming an asymmetric size distribution, the net effect would be a general red-shift and broadening of the resonance peak, especially for those resonances of low energy (long wavelengths) and associated to the metallic phase which are more sensitive to changes in the morphology and optical properties of the VO$_2$ nanoscatterers. 

Other effects can also present minor contributions to the discrepancies observed in Figure \ref{fig3}d-e. Indeed, we approximate VO$_2$-NCs as composed of purely stoichiometric VO$_2$ and we neglected the presence of suboxides such as V$_2$O$_5$ at the NCs surface. Furthermore, we directly used the dispersion as extracted from VO$_2$ thin films measurements to compute the different Mie coeffients. By doing so, we made the hypothesis that the complex permittivity of VO$_2$ inside the NCs has the exact same dispersion as VO$_2$ thin films, what could be subject to discussions~\cite{whittaker2011microscopic}.

Even though we have made all these approximations and neglected potential interactions in the fits, we find a remarkably good qualitative agreement between measurements and calculations. We emphasize that this simple model is able to reproduce well the prominent features of the optical response for both states of VO$_2$ by simply changing the dispersion of VO$_2$ from insulating to metallic. Indeed, the calculations nicely reproduce the first peak in the visible, which is a combination of electric and magnetic dipole resonances and the second peak in the NIR which is a plasmonic resonance. This latter is only present when the VO$_2$ is metallic, e.g. when the real part of the permittivity is negative and is therefore a dynamically tunable resonance.

These calculations enable us to retrieve the effective permittivity of the active VO$_2$-NCs layer and homogenize it as an effective metamaterial medium. 
 Figure \ref{fig4} a-d shows the calculated ${\epsilon_{eff}}$, ${\mu_{eff}}$ and $m_{eff}$ for a slab with VO$_2$ spheres with radius $R =$ 35 nm and a filling fraction $f =$ 0.18 in their insulating (low temperature) and metallic (high temperature) states.

We see clear differences in the optical dispersion as compared to un-patterned VO$_2$ layer. In particular, there are ranges of wavelength for which the IMT of VO$_2$ produces a large modulation of $n_{eff}$ with a simultaneously low $k_{eff}$ that is much lower than in bulk VO$_2$. This is especially true in the NIR, for which $k_{eff}$ remains lower than 0.1.

\vspace{-0.4cm}
\section{Discussion}
\vspace{-0.4cm}

\begin{figure*}[tb]
  \centering
  \includegraphics[width=1\linewidth]{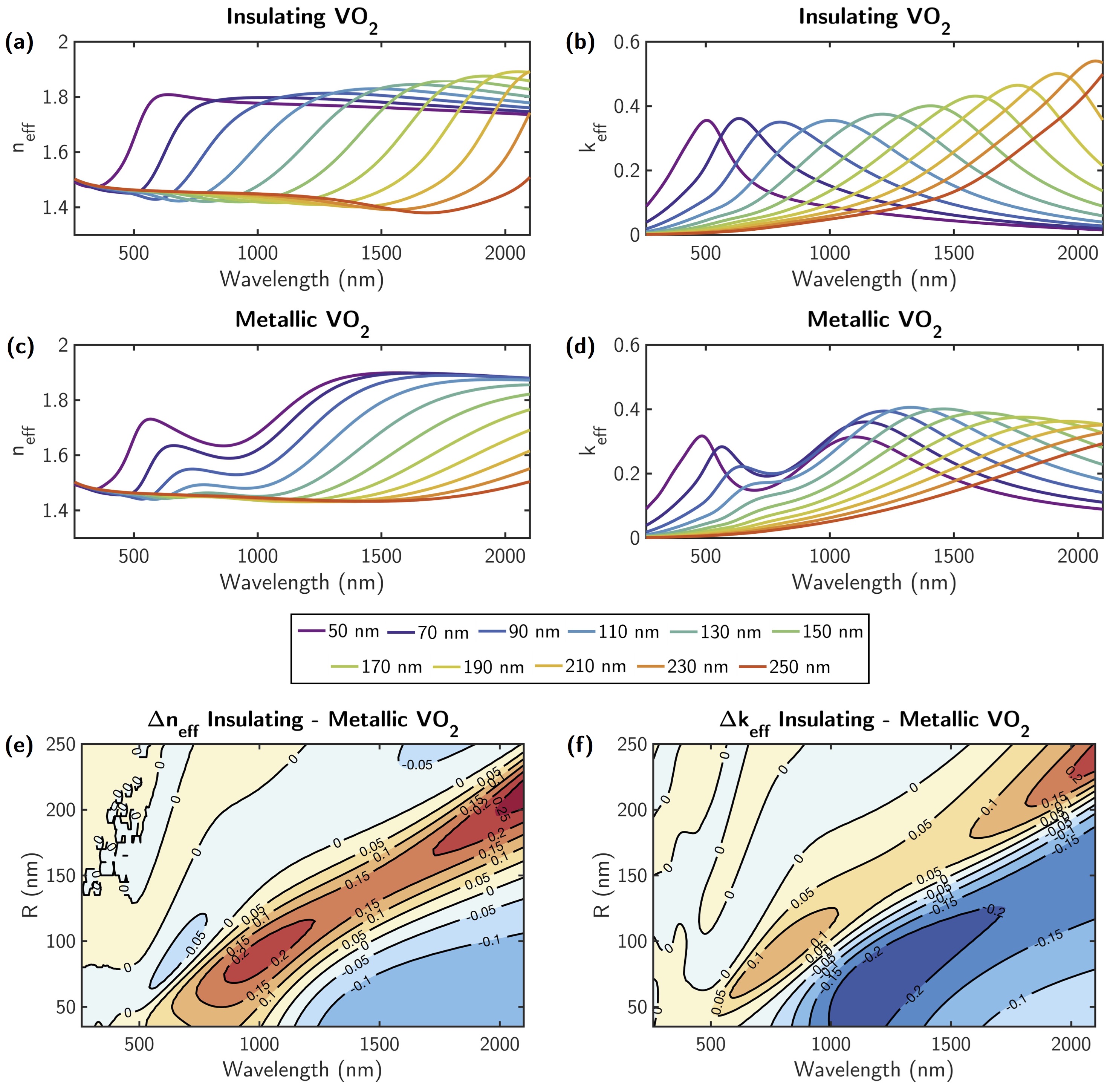}
  \caption{Variation of effective optical parameters for different sizes of nanocrystals in both states of VO$_2$  a) and b) n$_{eff}$ and k$_{eff}$ for insulating VO$_2$ with a filling fraction $f = $ 0.18. c) and d) n$_{eff}$ and k$_{eff}$ for metallic VO$_2$. e) and f) represent the amplitude of modulation of the effective refractive index of the metamaterial as a function of NCs size. e) $\Delta n_{eff}$ and f) $\Delta k_{eff}$.}
 \label{fig5}
\end{figure*}

We have shown that the main spectral characteristics of VO$_2$-NC based metamaterials can be calculated using a combination of Mie theory and Maxwell Garnett effective medium. From there, we can now predict the expected optical properties of VO$_2$-NCs of arbitrary sizes. 
In the following, we show how this framework can be used to design tunable metamaterials with tailored properties in desired wavelength ranges. Figure \ref{fig5}a-d shows the calculated evolution of the $n_{eff}$ and $k_{eff}$ values of VO$_2$-NCs in both their insulating and metallic states for different diameters of NCs.

In the insulating state, the absorption peaks are controlled by the Mie resonances while in the metallic state they are dominated by the plasmonic resonance. We can directly see that, by adjusting the size of NCs, we control the spectral position of the absorption peaks in both the insulating and metallic states (see figure \ref{fig5}b) and d).  The size of the VO$_2$-NCs therefore enables tailoring the spectral distribution of the multipolar resonances hosted by this metamaterial. 

\begin{figure}[tp]
  \centering
  \includegraphics[width=0.87\linewidth]{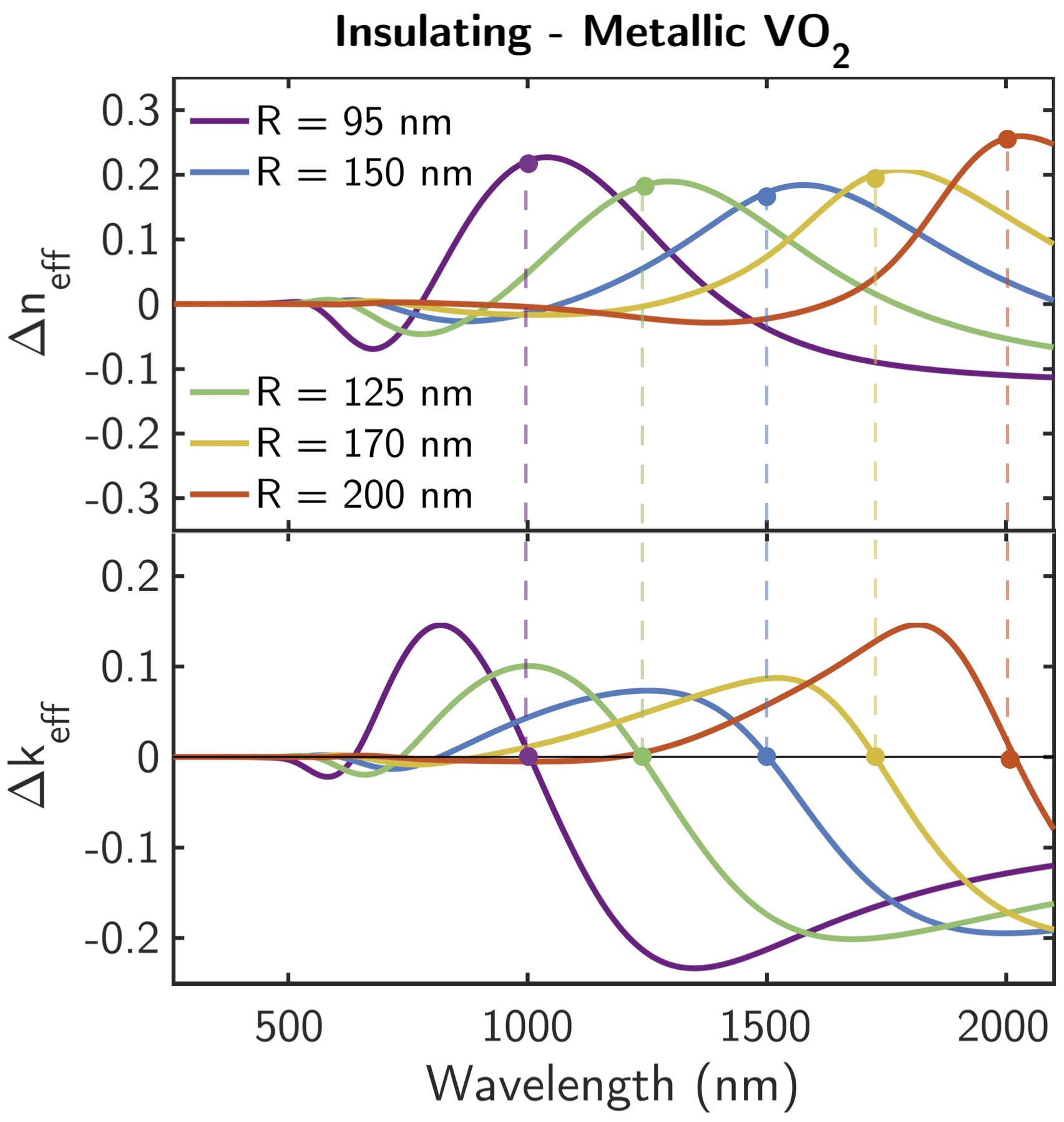}
  \caption{Spectral variation of the amplitude of modulation of the effective refractive index ($\Delta n_{eff}$ and $\Delta k_{eff}$) of the metamaterial for selected NCs size. }
  \label{fig6}
\end{figure}

To have a more general picture of the different regimes of tunability we can obtain with this system, Figure \ref{fig5}e-f displays the amplitude of refractive index modulation produced by the IMT of VO$_2$ as a function of NCs size. To do so, we plot $\Delta n_{eff}$ and $\Delta k_{eff}$, which are defined as $\Delta n_{eff}$=$n_{eff,insulating}$ - $n_{eff,metallic}$ and $\Delta k_{eff}$=$k_{eff,insulating}$ - $k_{eff,metallic}$ and which represent the amplitude of modulation of the effective refractive index and extinction coefficient, respectively.
With this figure, we can pinpoint specific regions of the size versus wavelength map in which the switching of VO$_2$ either translates into a tunable refractive index or a tunable extinction coefficient. 
Interestingly, by plotting the $\Delta n_{eff}$ and $\Delta k_{eff}$ values calculated for a few selected VO$_2$-NCs sizes (as displayed in Figure \ref{fig6}), we immediately see that around the maximum $\Delta n_{eff}$ reached for each NCs size, we simultaneously have $\Delta k_{eff}$=0. Furthermore, the wavelength at which this effect is obtained is directly governed by the size of VO$_2$-NCs. In other words, this phase-change metamaterial platform enables tailoring a dynamical tuning of refractive index without modifying the optical absorption. As displayed in Figure \ref{fig6}, this 'zero-induced-loss' refractive index tuning can be designed to occur throughout the NIR range by specifically selecting an appropriate size of NCs. This effect cannot be obtained with thin-films of phase-change materials and highlights the usefulness of modifying them through a metamaterial approach. We also note that the opposite effect can be exploited, namely, close to the maximum $\Delta k_{eff}$, we can simultaneously have negligible $\Delta n_{eff}$. So this platform can be used to actively control the absorption without modifying the refractive index of the medium (e.g. for R=95nm at $\lambda$=1500nm we get $\Delta k_{eff}$=-0.2 and $\Delta n_{eff}$=0).

\begin{figure*}[tb]
  \centering
  \includegraphics[width=\linewidth]{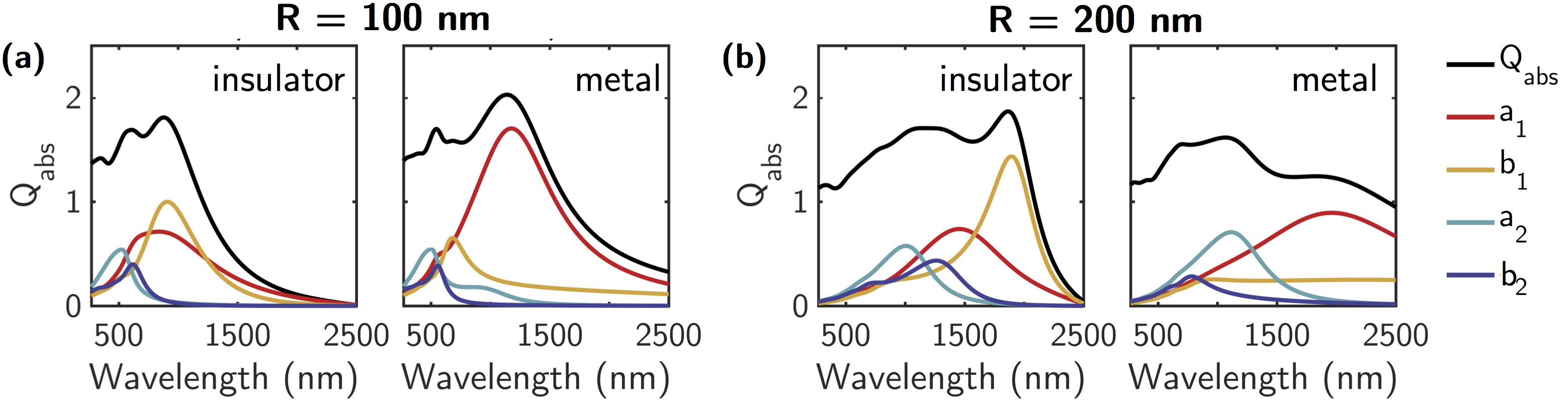}
  \caption{Calculated absorption cross-section efficiency $Q_{abs}$ of NCs of different sizes in their insulator and metallic states embedded in quartz. With colored lines are represented the dipolar electric $a_1$ (red), dipolar magnetic $b_1$ (yellow), quadrupolar electric $a_2$ (light blue) and quadrupolar magnetic $b_2$ (dark blue) contributions to $Q_{abs}$.}
  \label{fig7}
\end{figure*}

We envision another exciting opportunity for this metamaterial as an active platform to control scattering properties and directional beaming of light. Indeed, when multipolar resonances are spectrally overlapping in nanoparticles, the scattered fields can interfere and produce directional scattering. In particular, spectrally overlapping DE and DM resonances yield the so-called Kerker conditions that either suppress the forward or the backward scattering of light~\cite{kerker1983electromagnetic,Geffrin2012}. Figure \ref{fig7} displays the multipolar decomposition of light fields within VO$_2$-NCs as a function of size and VO$_2$ phase, either insulating or metallic. We see that each size of NCs produces overlapping DE and DM resonances which coherently interfere at specific wavelengths. Then, when the VO$_2$ is switched  from its insulating phase to the metallic one, we can suppress the  DM contribution associated to the $b_{1}$ coefficient and strongly enhance the DE contribution associated to the $a_{1}$ coefficient. In other words, at specific wavelengths we switch from a dielectric DM resonances to a plasmonic DE resonance. This implies that this hybrid metamaterial enables an active and dynamical control of the Kerker conditions~\cite{kerker1983electromagnetic}. From that perspective, we foresee a dynamic modulation of the directionality of light scattering as well as actively tunable perfect reflection or absorption. We also note the presence of tunable quadrupolar modes ($a_{2}$ and $b_{2}$ coefficients, figure \ref{fig7}) that may be used for complex tunable interference phenomena between different multipoles, hence opening a wealth of different possibilities of spatially distributing the scattered light for directionality control purposes (see~\cite{Tribelsky2015, liu} for more details). This tunable multipoles could also find an interest in the recently proposed framework of spontaneous emission engineering through intereferences between higher-order multipoles~\cite{rusak}.

\vspace{-0.4cm}
\section{Conclusion}
\vspace{-0.4cm}

We have demonstrated that VO$_2$-NCs integrated in SiO$_2$ as spherical nano-inclusions enable exploiting the VO$_2$ IMT to tune and switch multipolar modes in the visible and near-infrared ranges. In particular, we showed the presence of a plasmonic mode in the NIR whose gradual appearance and intensity are directly controlled by the VO$_2$ state. The presence of two simultaneous dipole resonances in two different spectral regions, Vis and NIR, could open the possibility of new “multiplexing” based techniques for SERS applications. Contrary to bulk and thin-films VO$_2$, such a metamaterial approach provides means to tailor the complex refractive index and tunability of phase-change materials. Indeed, we have shown that, by adjusting the VO$_2$-NCs size, this system can be exploited as a designer phase-change metamaterial with adjustable refractive index modulation and absorption modulation. Especially, we can design the system to have a large refractive index tunability without inducing modulation of the extinction coefficient at specific wavelengths, or in other words a zero-induced-loss refractive index modulation. 
Furthermore, such VO$_2$-NCs can be optically switched at femtoseconds timescale ($\sim$150 fs)~\cite{rini} and present very large hysteresis (see ~\cite{jostmeier} and the supplementary information). We can therefore envision this platform to be used for ultra-fast all-optical integrated systems with memory effects.

We emphasize that the resulting metamaterial presents several benefits from a practical point of view, and most notably the nanostructures are embedded in a dielectric host that protects them from the environment and consequent processes that can affect its optical response (protection from oxidization \cite{Gutierrez2016}, chemical reaction \cite{Oates2013}, mechanical scratches,...). This makes such a platform extremely easy to handle and to integrate in various environments, contrary to thin-films of phase-change materials that are very sensitive. 
The samples presented here were fabricated by ion implantation, a technique that is widely used in the microelectronics industry and could therefore be exploited for large-scale fabrication of samples. Furthermore, the size, density and depth of NCs in the host could be controlled through adjustment of implantation parameters and annealing process. This implantation process can be carried out in many different CMOS-compatible platforms such as thermal SiO$_2$ on silicon substrate and results in a flat surface that does not require any complex post-process step such as CMP to planarize the device. A next logical step could be to integrate nanophotonic devices on top of this metamaterial for a wealth of tunable functionalities such as an active control of spontaneous light emission~\cite{bidault2019dielectric,rusak,cueff2019tailoring} or dynamically tunable Huygen's metasurfaces~\cite{decker2015} .

From a broader scientific perspective, the results presented here help define a new system with which to study tunable interactions between nanoparticles and tunable interferences between multipolar resonances. We envision it to serve as a testbed for dynamical modulation of light scattering, tunable Kerker effects and active control of light directionality. 

\vspace{-0.4cm}
\section{Experimental section}
Fabrication of VO$_2$ thin-films: 25 nm VO$_2$ was deposited on quartz substrate by magnetron sputtering. A V$_2$O$_5$ target was sputtered at a power of 100 W radio frequency. During deposition, the substrate was maintained at 750$^\circ$C; and the pressure of chamber was kept as 5 mtorr with flowing of 49.5 sccm Ar and 0.5 sccm 10\% Ar-balanced oxygen gases. 

The synthesis of a dense layer of isolated VO$_2$ nanocrystals embedded in about 100 nm depth below the surface begins with a high fluence ion implantation of V$^{+}$ and O$^{+}$ with energies of 100 keV and 36 keV, respectively. These energies were chosen to produce overlapping concentration depth profiles of the elements. The formation of the VO$_2$ chemical compound was controlled by the fluence ratio of V to O (i.e. 8$\times$10$^{16}$ cm$^{-2}$ to 1.6$\times$10$^{17}$ cm$^{-2}$ and 4$\times$10$^{16}$ cm$^{-2}$ to 8$\times$10$^{16}$ cm$^{-2}$ for a ratio of 1:2 ). After ion implantation nanocrystal growth was initiated by a rapid thermal annealing (RTA) step at $1000^\circ$C for 10 min in an inert gas at atmospheric pressure.

Measurements of the complex dielectric functions of all samples were carried out using a variable-angle spectroscopic ellipsometer (UVISEL Horiba Jobin-Yvon). The incident broadband light source (Xenon) was polarized at 45$^\circ$. After reflection off the sample under study, the light is directed to a spectrometer, dispersed by appropriate gratings and measured with UV/Vis and NIR detectors. The acquired ellipsometric parameters $Is$ and $Ic$ of the samples have been collected for varying angles 55$^\circ$ - 75$^\circ$ over a spectral range of 260 - 2100 nm. The complex dielectric functions of samples and their thicknesses can be derived by fitting realistic optical models to the experimental data (more details in the supplementary information). 
To analyze temperature-dependent optical properties, the acquisitions were carried out in a range of temperatures from 30$^\circ$C to 80$^\circ$C using a digitally controlled heat cell (Linkam THMSEL350V). 

\vspace{-0.4cm}
\begin{acknowledgments}
\vspace{-0.4cm}
This work is partly supported by the French National Research Agency (ANR) under the project SNAPSHOT (ANR-16-CE24-0004).
Y.G. and F.M. acknowledge the support by the Army Research Laboratory under Cooperative Agreement Number W911NF-17-2-0023 and by SODERCAN (Sociedad para el Desarrollo de Cantabria) through the Research Vicerrectorate of the University of Cantabria. Y.G. thanks the University of Cantabria for her FPU grant. ZZ and SR acknowledge AFOSR FA9550-18-1-0250 for support. The authors thank Prof. David E. Aspnes for helpful discussions.
\end{acknowledgments}

\bibliography{Reference}

\end{document}